# Bias-tunable two-dimensional magnetic and topological materials

Jie Li and Ruqian Wu*

Department of Physics and Astronomy, University of California, Irvine, California 92697-4575, USA.

Searching for novel two-dimensional (2D) materials is crucial for the development of the next generation technologies such as electronics, optoelectronics, electrochemistry and biomedicine. In this work, we designed a series of 2D materials based on endohedral fullerenes, and revealed that many of them integrate different functions in a single system, such as ferroelectricity with large electric dipole moments, multiple magnetic phases with both strong magnetic anisotropy and high Curie temperature, quantum spin Hall effect or quantum anomalous Hall effect with robust topologically protected edge states. We further proposed a new style topological field-effect transistor. These findings provide a strategy of using fullerenes as building blocks for the synthesis of novel 2D materials which can be easily controlled with a local electric field.

* E-mail: wur@uci.edu.

# 1. INTRODUCTION

The discovery of excellent functional materials is the foundation for major technological innovations, and this has been manifested again in the new wave of research interest in diverse two-dimensional (2D) materials. Ever since the successful exfoliation and synthesis of graphene,[1] a great deal of efforts have been dedicated to fabricate and manipulate new 2D materials, from semiconducting transition metal dichalcogenides[2,3] 2D van der Waals (vdW) magnetic monolayers,[4,5] to 2D topological insulators[6,7]. Due to their unique physical properties, many 2D materials are promising for the design of the next generation electronics, optoelectronics, electrochemistry and biomedicine devices.[8-13] During the course of the development of emergent functional 2D materials, theoretical predictions via density functional theory (DFT) calculations have played a major role by predicting possible material combinations and providing insights for experimental observations. They inspired tremendous experimental interest and a decent number of hypothetical materials have been actually synthesized in experiments. Remarkable examples in this category include metal-organic frameworks (MOFs),[14-18] covalent-organic frameworks (COFs),[19-21] graphene-based systems,[3, 22-24] thin metal nanostructures,[25,26] and Mxenes,[27,28] to name a few. Obviously, reliable theoretical predictions for new functional material systems are crucial to propel this exciting field forward.

Endohedral fullerenes with metal atoms, ions, or clusters embedded in carbon cages can give rise to exotic properties by adjusting their ingredients, size and symmetry. Superconductivity,[29,30] ferroelectricity,[31-33] and single-molecule magnetism[34-36] have been reported for these systems. Buoyed by the ongoing effort of the synthesis of 2D fullerenes on graphene or transition metal surfaces by self-assembly process,[37-39] we proposed that W@$C_{28}$ may serve as building blocks for the synthesis of multifunctional 2D materials in our previous work.[40] W@$C_{28}$ has structural bistability, ferroelectricity, multiple magnetic phases, and excellent valley characteristics. Furthermore, it can be functionalized as a valleytronic material by integrating with magnetic insulators such as the $MnTiO_3$ substrate or by switching a small amount of them to the magnetic metal stable state. This inspires us to examine $C_{28}$ fullerenes with different 3d transition metal insertions and search for their potentially useful properties.

In this paper, we construct various 2D materials with M@$C_{28}$ endohedral fullerenes (M=3d transition metals) as building blocks. Through systematic ab-initio calculations and model simulations, we find that many of them are ferroelectric, with the core metal atoms taking two stable sites. Some of them simultaneously have multiple magnetic phases with large magnetic anisotropy energies and high Curie temperatures, ideal for magnetoelectric operations. Furthermore, Ti and Cr cores in the honeycomb lattice produce standalone 2D topological insulators, with robust quantum spin Hall effect (QSHE) and quantum anomalous Hall effect (QAHE), respectively. These results indicate that 3d M@$C_{28}$ endohedral fullerenes may form excellent 2D multifunctional materials for technological innovations. To show the possible application of their unique physical properties, we further propose a design of topological field-effect transistor.

## 2. METHODOLOGY

All density functional theory calculations in this work are carried out with the Vienna ab-initio simulation package (VASP) at the level of the spin-polarized generalized-gradient approximation (GGA) with the functional developed by Perdew-Burke-Ernzerhof.[41] The interaction between valence electrons and ionic cores is considered within the framework of the projector augmented wave (PAW) method.[42,43] The energy cutoff for the plane wave basis expansion is set to 500 eV. All atoms are fully relaxed using the conjugated gradient method for the energy minimization until the force on each atom becomes smaller than 0.01 eV/Å, and the criterion for total energy convergence is set at $10^{-5}$ eV. The one-dimensional (1D) band of nanoribbon is calculated with a tight-binding (TB) model based on the maximally localized Wannier functions (MLWFs), as implemented in the Wannier90 code.[44]

## 3. RESULTS AND DISCUSSION

As many endohedral fullerenes have been successfully synthesized,[29-36,45] it is attractive to use them as building blocks for the construction of 2D materials. In this work, we focus on endohedral M@$C_{28}$, M=Ti-Zn that are the smallest stable endohedral fullerenes synthesized in carbon vapor.[45] Typical single endohedral $C_{28}$ molecules have inherent electric dipoles as the core atoms shift away from the center of the carbon cage, as sketched for the up- and down polarizations in Fig. 1(a), denoted as phase I and phase II below. They also have the $C_{3v}$ symmetry about the vertical axis, and we hence perceive that highly symmetric 2D covalent crystals may form in either the close-packed or honeycomb lattice as shown in Fig. 1(b) and 1(c), respectively. As a result, each M@$C_{28}$ has six possible lattice-polarization combinations. As shown by the relative energies with respect to the close-packed M@$C_u$ case (set as zero) in Fig. 1 (d), ab-initio calculations indicate that Ti (V, Cr, Fe)@$C_{28}$ prefer the close-packed structure, while others prefer the honeycomb lattice in their ground state. Nevertheless, energy differences among these three lattices are not large for most fullerenes (< ~0.5 eV per molecule) except for the Ti cases. The energy differences for several Cr cases ($C_u$, $C_d$, $H_{1-u}$ and $H_{2-u}$) are even smaller than 0.1 eV. We may assume that all of them can be synthesized in experiments, depending on the substrate and growth conditions. Fig. S2 shows that the honeycomb structure overtakes the close-packed one for Cr@C$_{28}$ as the density of fullerenes is reduced. To keep the following discussions more general, we mostly give properties of these 2D materials in all three lattices below.

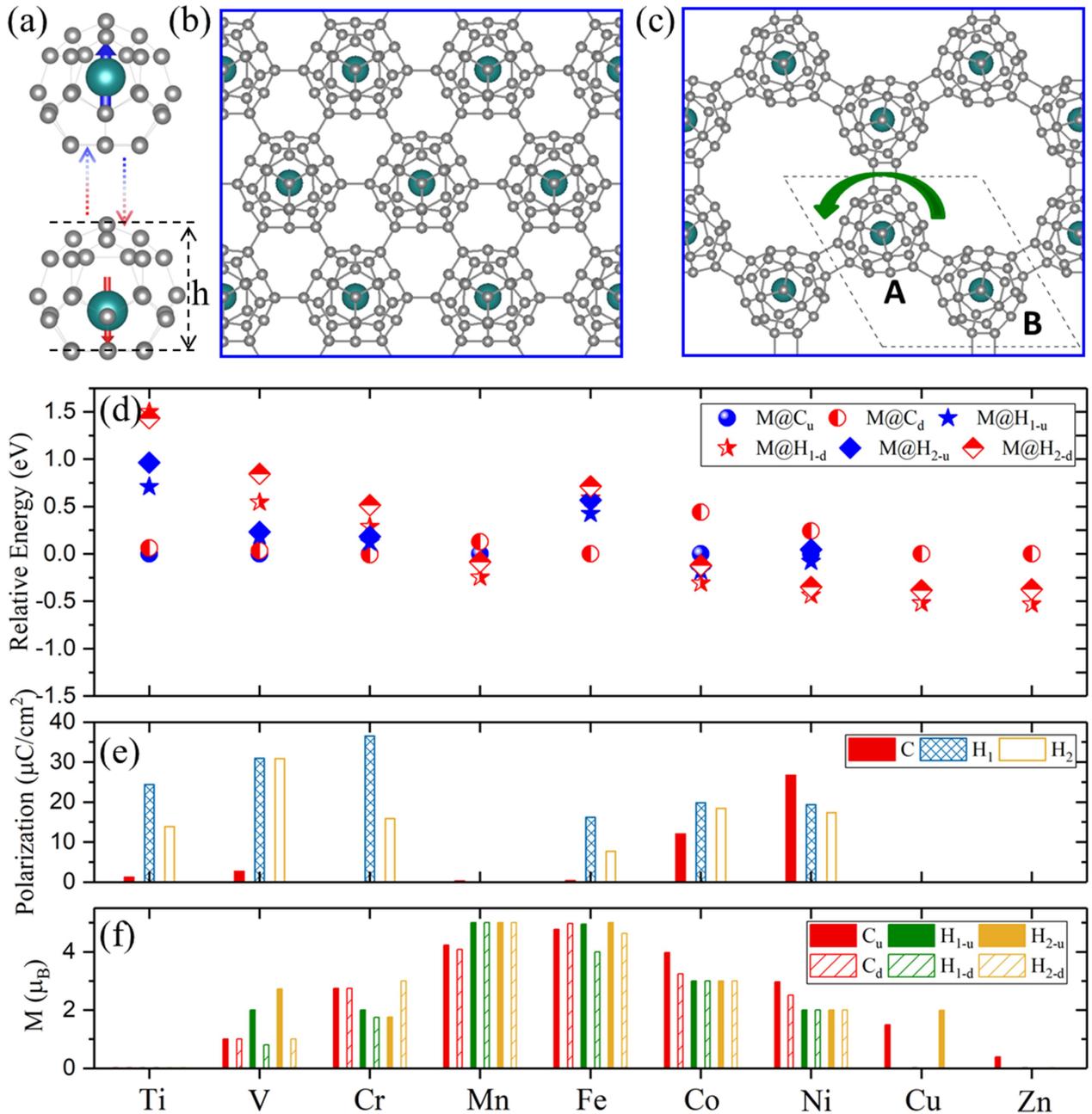

**Fig. 1** (a) Schematic structures of M@C$_{28}$ in phases I and II, two colored arrows represent the

direction of their dipoles. 2D fullerene crystals in (b) close-packed lattice (M@C) and (c) honeycomb lattice with either the $C_{3v}$ symmetry (M@H$_1$) or $C_3$ symmetry in which the molecule at the A site rotates by ~62° (M@H$_2$). (d) The relative total energies of 2D endohedral $C_{28}$ with respect to that of M@C$_u$. (e) The difference of electric dipole moment between the up- and down polarizations. (f) The magnetic moments per molecule in different lattice-polarization combinations.

The structural stability of these lattices is indicated by their binding energies in Table S1, which is defined as $E_b = E_{mol} - E_{2D}/n$. Here, $E_{mol}$ and $E_{2D}$ are the total energies of the isolated M@C$_{28}$ molecule and the 2D crystal; and $n$ is the number of M@C$_{28}$ molecules per unit cell. Large positive values of $E_b$ suggest strong interactions among M@C$_{28}$ molecules, helpful for synthesizing M@C$_{28}$ 2D covalent crystals in experiments. Furthermore, the phonon calculations and ab initio molecular dynamics (AIMD) simulations are used to check their thermal and dynamic stability. Here, we take Cr@C$_u$, Cr@H$_{1-u}$ and Cr@H$_{2-u}$ as examples. The corresponding phonon dispersions in Fig. S1 show the absence of imaginary frequency branch, indicating that these systems are dynamically stable. In the AIMD simulations with a 6×6 (4×4) supercell of 1044 (928) atoms for Cr@C$_u$ (Cr@H$_{1-u/2-u}$), the structures are not noticeably deformed after 5000 steps (10ps) at 300K. The total energies fluctuate around their equilibrium values without sudden jump as shown in Fig. S1. Therefore, we believe that these 2D covalent crystals are thermally stable at least up to room temperature.

Interestingly, we find that almost all M@C$_{28}$ 2D crystals have two stable structural phases as depicted in Fig. 1 (a), except for Mn, Cu, Zn which has no unpaired $d$ electrons in the carbon cage. By switching between their two structural phases, the metal cores shift along the vertical direction and the corresponding atomic displacements are list in Table S2. Nevertheless, the carbon cage noticeably deforms while forming the close-packed lattice, as suggested by the decrease of height (cf. data in Table S3 for h) and hence the displacements are small for M@C geometries, except Co and Ni which have more localized d-orbitals and smaller size. As listed in Table S4, most energy barriers between these two stable structural phases are smaller than 0.5 eV, so the phase change can be driven by bias. As metal cores donate electrons to the carbon cage, there are net electric dipoles in these molecules, and the 2D lattice may manifest strong ferroelectricity, as we discussed before for W@C$_{28}$.[37] Quantitatively, the electric dipole moments are calculated. The results are shown in Fig. 1(e). Taking Cr@H$_1$ as an example, the Bader charge analysis shows that each Cr atom donates 1.26 e (1.21e) to the carbon cage in phase I (II) (see charge redistribution in Fig. S3), and the Cr atom shifts down by 1.22Å as switching from phase I to phase II which have dipole moments of -0.268 eÅ and 0.292 eÅ per molecule, respectively. These large and opposite electric dipole moments indicate that Cr@H$_1$ has exceedingly strong ferroelectric polarization (36.5μC cm$^{-2}$) compared to other 2D ferroelectric materials (e.g., 1.6μC cm$^{-2}$ for Sc$_2$CO$_2$, 18.1μC cm$^{-2}$ for SnSe, and 19.4μC cm$^{-2}$ for SnTe).[46-48]

Meanwhile, DFT calculations reveal that most of these fullerene lattices have large local magnetic moments in Fig. 1(f) and also ferromagnetic couplings (see in Fig. S4). The corresponding exchange parameters (see in Table S5) are obtained by mapping the DFT total energies of different magnetic configurations to the classical Heisenberg Hamiltonian:

$$H = H_0 - J_1 \sum_{<i,j>} S_i \cdot S_j \qquad (2)$$

where $J_1$ represent the nearest neighbor exchange interactions. We only consider $J_1$ as the separations between fullerene molecules are large. To determine the thermal stability of their magnetic phase, we calculate their magnetic anisotropy energies (MAEs) by using the torque method (see in Fig. S4).[49,50] One may see that Cr, Mn, Fe, Co and Ni cases have large MAEs, especially for Fe@H$_{2-d}$ (1.02meV per molecule), Co@C$_d$ (-2.47meV per molecule) and Ni@C$_u$ (-2.36meV per molecule). With the exchange parameters in Table S5 and magnetic anisotropy energies in Fig. S4(a), we further calculate Curie temperatures ($T_c$) of these new 2D magnetic materials by using the renormalized spin-wave theory (RSWT)[51,52] and results are shown in Fig. S4(b). Assuming the Curie temperature as the renormalized magnetization drops zero, we find that several these systems have comparably higher Curie temperatures than many existing 2D magnetic materials, such as CrI$_3$ (45K) and Cr$_2$Ge$_2$Te$_6$ (66K).[53,54] Especially, Fe@H$_{1-d}$ is half-metallic (cf. the DOS curves Fig. S5) and has a Curie temperature as high as 147K, which makes it very attractive as an efficient spin filter for spintronic applications.

It is intriguing to investigate if these systems may have nontrivial topological properties for the realization of quantum spin Hall effect and quantum anomalous Hall effect which may find applications in quantum information technologies. Through systematic search from their band structures, we find that the nonmagnetic Ti@H$_{1-u}$ has a SOC-induced band gap of 10.8 meV right at the Fermi level, a desirable feature for topological materials (see in Fig. 2(a)). To verify if this gap is topologically nontrivial, the Z$_2$ number of Ti@H$_{1-u}$ is calculated. In Fig. 2b, one may see that Z$_2$=1 for Ti@H$_{1-u}$ by counting the positive and negative n-field numbers over half of the torus. This undoubtedly indicates that Ti@H$_{1-u}$ is a new 2D topological insulator that may manifest the QSHE.As a further evidence, we construct a zigzag Ti@H$_{1-u}$ nanoribbon about 30nm in width (see in Fig. 2(c)) and check if it has the topologically protected edge states. Due to large number of atoms in the nanoribbon, we build a TB model with parameters obtained by fitting the DFT band structure of 2D Ti@H$_{1-u}$. Using *s* and *p* orbitals for C and *s* and *d* orbitals for Ti in the bases, this TB model gives excellent band structures near the Fermi level comparing with the DFT results (see in Fig. S6). In Fig. 2(d), one may see two edge bands intercepting the Fermi level and connecting the bulk conduction and valence bands in the two sides of the Γ point. This further supports that Ti@$H_{1-u}$ is a new 2D topological insulator and may offer a possible platform for the realization of the quantum spin Hall effect. In contrast, the Ti@$H_{1-d}$ lattice is a trivial insulator, and the topological phase transition therefore can be controlled by the ferroelectric polarization.

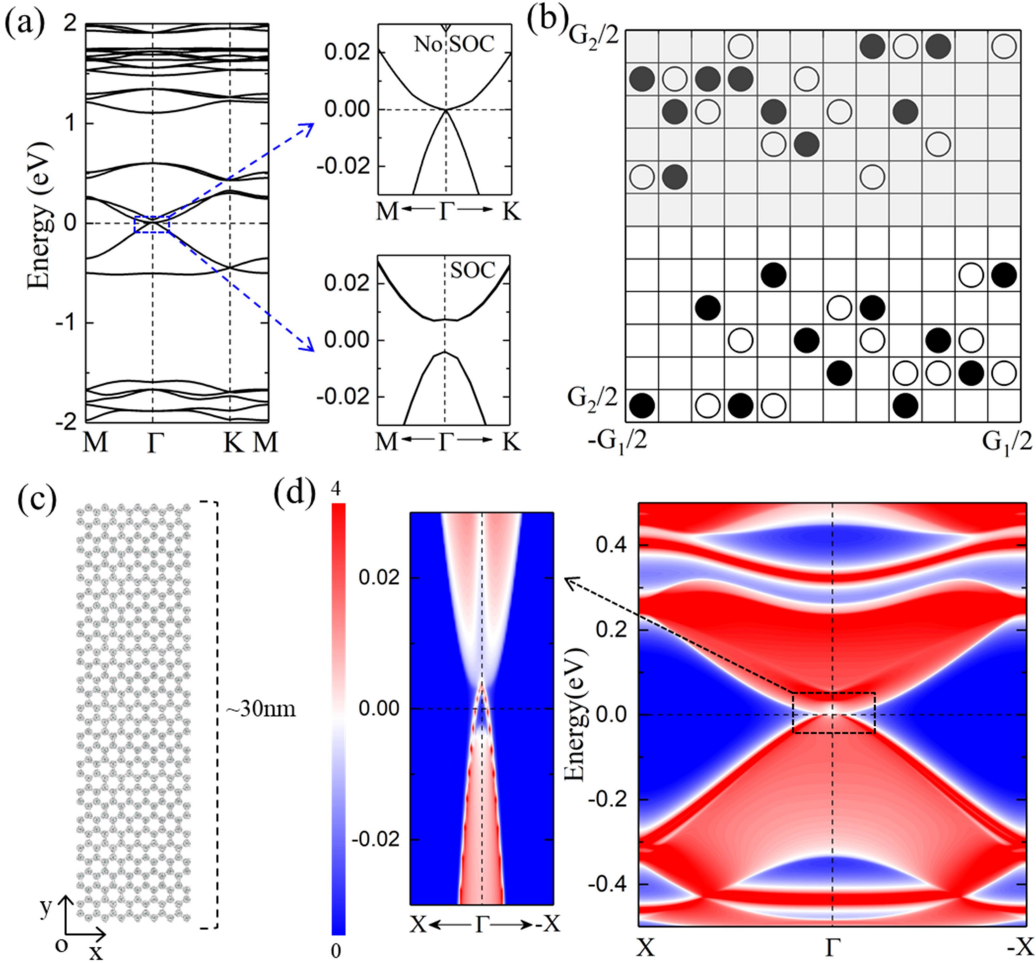

**Fig. 2** (a) The band structures of Ti@$H_{1-u}$ without and with SOC. (b) The n-field configuration with solid, hollow circles and blank boxes denoting n= -1, n= 1 and n= 0. (c) The schematic structure of a zigzag Ti@$H_{1-u}$ nanoribbon about 30nm in width (along y-axis) and keeps the periodicity along the x-axis. (d) The corresponding 1D band structure and the top edge states.

As many fullerene lattices are magnetic, it is intriguing to examine if they may manifest the QAHE. Due to the difficulty of controlling the distribution and magnetic order of dopants or the strong interfacial hybridization, very few successful observations of QAHE have been reported using the conventional approaches that magnetize the topological surface states with doping or interfacial proximity.[55-57] Here, our calculations show that Cr@H$_{1-u}$ has a SOC-induced spin polarized band gap of 7.4 meV right at the Fermi level (see in Fig. 3(a)), and its local magnetic moment is 2$\mu_B$ per molecule. To verify if Cr@H$_{1-u}$ is a Chern insulator, its Berry curvature $\Omega(k)$ is calculated in the Brillouin zone as

$$\Omega(k) = 2\text{Im} \sum_{n\epsilon\{o\}} \sum_{m\epsilon\{u\}} \frac{\langle\psi_{nk}|v_x|\psi_{mk}\rangle\langle\psi_{mk}|v_y|\psi_{nk}\rangle}{(\varepsilon_{mk}-\varepsilon_{nk})^2} \quad (4)$$

Here, $\{o\}$ and $\{u\}$ are the sets of occupied and unoccupied states; $\psi_{nk}$ and $\varepsilon_{nk}$ are the Bloch wave function and eigenvalue of the *n*th band at the k point; and $v_x$ and $v_y$ are the velocity operators, respectively. One may see that large positive Berry curvature presents around the Γ point, indicating that the gap is truly topologically nontrivial (see in Fig. 3(b)). Meanwhile, the Chern number (C), which gives the Hall conductance as $\sigma_{xy} = C(e^2/\hbar)$, is directly calculated by integrating the Berry curvature in the Brillouin zone as

$$C = \frac{1}{2\pi} \int_{BZ} \Omega(k) d^2 k \quad (5)$$

In Fig. 3(c), one may see that a small terrace with *C* = 1 presents in the SOC induced band gap, which is the hallmark of the QAHE materials. As further evidence, the top and bottom edge states of a zigzag Cr@H$_{1-u}$ nanoribbon can be seen in the bulk gap according to the TB calculations with parameters extracted from the DFT bands. The top edge state moves along the positive x-axis while the bottom edge state moves along the opposite direction (see in Fig. 3(d)). Obviously, a net

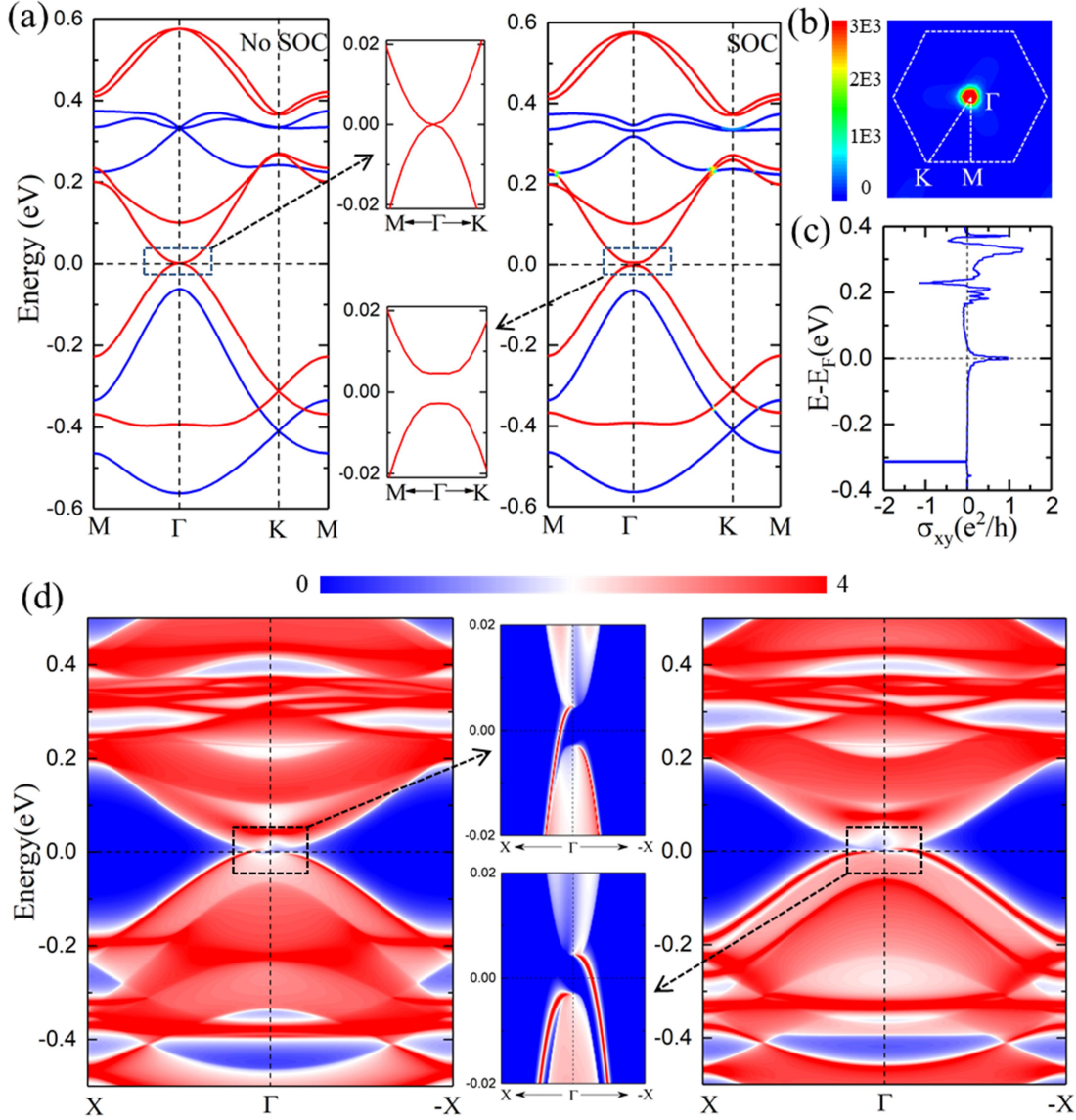

**Fig. 3** (a) The band structures of Cr@$H_{1-u}$ without and with SOC (the middle are the corresponding zoom-in band structures near the Fermi level around the Γ point). (b) and (c) The distribution of Berry curvature in the 2D Brillouin zone and Fermi level-dependent anomalous Hall conductance ($\sigma_{xy}$) for Cr@$H_{1-u}$, respectively. (d) The corresponding 1D band structure and the top and bottom edge states.

quantized transverse Hall current can be expected when zigzag Cr@$H_{1-u}$ nanoribbons are used in nanodevice. Since this material has a Tc of 14.3 K, we may expect that the QAHE can be observed with the liquid Helium4 temperature, which is attainable in most labs. The discovery of these standalone 2D topological insulators is exciting as they may avoid technical complexities in dealing with doping and unwanted interfacial hybridization with the conventional approaches. We believe it is particularly rewarding for experimentalists to try synthesizing Ti@$C_{28}$ and Cr@$C_{28}$ honeycomb lattices and using their topological properties.

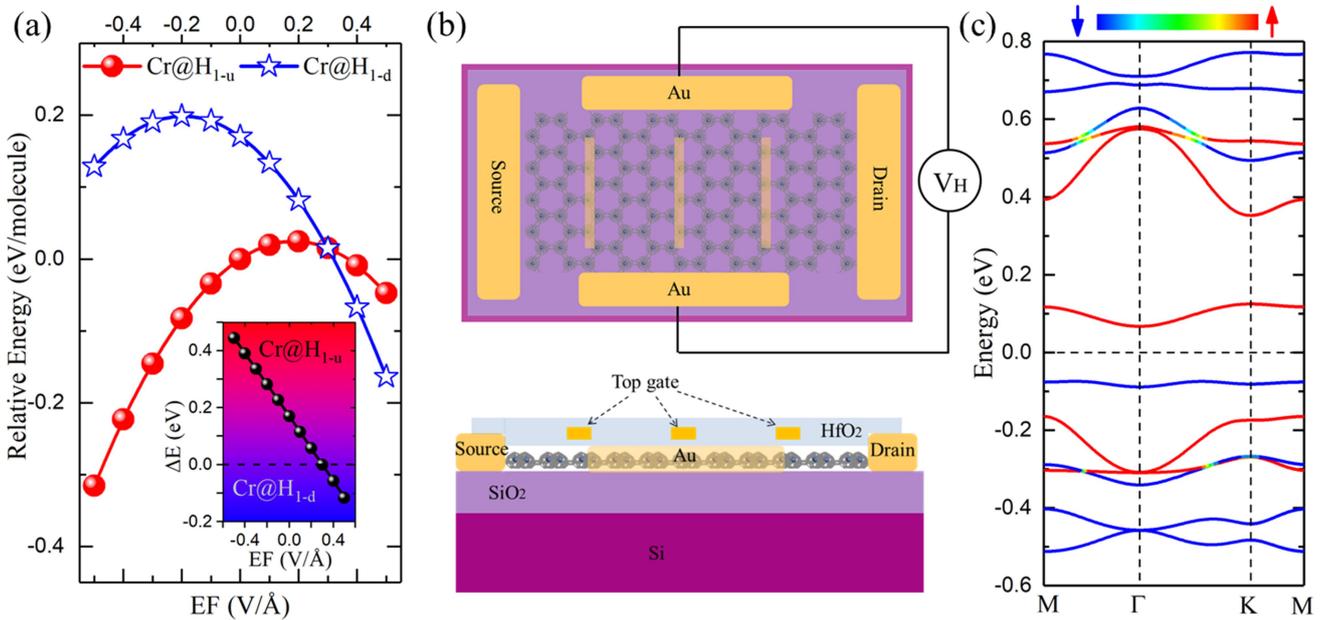

**Fig. 4** (a) The relative energy of Cr@$H_{1-u/d}$ as functions of the external electric field. (b) Schematic topological field-effect transistor based on Cr@$H_{1-u/d}$, (top and side view). Two Au electrodes are for the detection of Hall voltage. The nano-electrode buried in $HfO_2$ are for gate control. (c) The band structure of Cr@$H_{1-u}$ with a phase transition concentration of 50%.

Finally, we discuss how to combine the topological and ferroelectric properties of Cr@$H_{1-u}$ for the design of spintronic devices. By using a gate bias, we may realize a reversible structural phase transition as shown in Fig. 4(a), as also drastically change the electronic and magnetic properties of the M@$C_{28}$ lattice. This interplay allows us to design a conceptual topological field-effect transistor as shown in Fig. 4(b). We use a doped silicon/SiO$_2$ film as the substrate and four Au electrodes for the measurement of transvers voltage and longitudinal current. An ultrathin HfO$_2$ (or hexagonal BN) overlayer provides a protective environment, with nano-electrodes buried underneath as the top gate. As discuss above, the Cr@H$_{1-u}$ nanoribbon manifests the QAHE at a reasonably low temperature, and hence have spin polarized current at the edges. When some Cr@C$_{28}$ molecules are converted to the phase II, the ribbon switches to an ordinary magnetic semiconductor, as shown by the bands in Fig. 4(c) and Fig. S7. The transport properties are thus highly tunable by the top gate (~100%) is this geometry, which behaves as an ideal topological field-effect transistor. With such an easy controllability and unique topological feature, it is perceivable that many different devices can be designed with these new 2D fullerene lattices.

## 4. Conclusion

In summary, by using systematic ab-initio calculations and model simulations, we found several important new 2D functional materials based with 3d M@C$_{28}$ molecules in hexagonal or honeycomb lattice. Explicitly, 1) Fe@C$_{28}$ honeycomb lattice is a halfmetallic materials with a Tc up to 147 k; 2) Ti@C$_{28}$ honeycomb lattice is a topological insulator with a gap for 10.8 meV right across the Fermi level; and 3) Cr@C$_{28}$ honeycomb lattice is a Chern insulator with a gap of 7.4 meV. The discovery of these standalone functional materials may allow the realization of 100% spin filtering, quantum spin Hall effect or quantum anomalous Hall effect, each of them is an important research subject for the development of spintronics and quantum information technologies. Another advantage of these 2D materials is the easy controllability via electric field as most of them are intrinsically ferroelectric. As an example, we demonstrated a conceptual topological field-effect transistor based on the Cr@H$_{1-u}$ lattice, which can be switched from Chern insulator to ordinary magnetic semiconductor by tuning some Cr@C$_{28}$ molecules to metastable structural phase II with a local electric field. This work enriches the family of 2D magnetic and topological materials and points to a way for integrate different functionalities in a single simple system for diverse technological innovations.

## Author contributions

J.L. and R.W. designed the studies. R.W. conceived this project. J.L. performed the calculations and analyzed data. Both authors prepared the manuscript.

## Conflicts of interest

The authors declare no competing financial interest.

## Acknowledgments

Work is supported by US DOE, Basic Energy Science (Grant No. DE-FG02-05ER46237). Calculations are performed on parallel computers at NERSC.